\documentstyle[12pt,epsfig]{article}           
\newcommand{\be}{\begin{equation}}              
\newcommand{\ee}{\end{equation}}              
\newcommand{\bear}{\begin{eqnarray}}              
\newcommand{\eear}{\end{eqnarray}}              
\newcommand{\ba}{\begin{array}}              
\newcommand{\ea}{\end{array}}

\newskip\humongous \humongous=0pt plus 1000pt minus 1000pt

\newif\ifdtup

\def\oldreffmt#1{\rlap{[#1]} \hbox to 2\parindent{}}

\def\figfmt#1{\rlap{Figure {#1}} \hbox to 1in{}}              
              
\def\slash#1{#1\!\!\!/\!\,\,}              
\def\beq{\begin{equation}}              
\def\eeq{\end{equation}}              
\def\bea{\begin{eqnarray}}              
\def\eea{\end{eqnarray}}              
\def\half{\frac{1}{2}}              
              
\def\bq{\begin{quote}}              
\def\eq{\end{quote}}

\def\half{\frac{1}{2}}                   
\relax              

\newdimen\tdim              
\tdim=\unitlength              
\def\bar{\overline}

\begin{document}              
              
\pagestyle{empty}              
\begin{titlepage}              
\def\thepage {}    % Kill page numbering              
         \title{   \vspace*{1.5cm}  \bf                
Deconstructing $5$-D QED }               
\author{  \\ [0.5cm]           
\bf  Christopher T. Hill \\[2mm]              
\bf Adam K. Leibovich \\ [2mm]              
{\small {\it Fermi National Accelerator Laboratory}}\\            
{\small {\it P.O. Box 500, Batavia, Illinois 60510-0500, USA}}            
\thanks{            
hill@fnal.gov, adam@fnal.gov }\\            
}            
\maketitle            
\vspace*{-10.0cm}            
\noindent            
\begin{flushright}              
FERMILAB-Pub-02/078-T \\ [1mm]              
May, 2002              
\end{flushright}

%\ \vspace*{14.1cm}              
\vspace*{12.1cm}              
\baselineskip=18pt              
              
\begin{abstract}              
              
  {\normalsize              
We discuss periodic compactification      
and latticization of a  $5$-D $U(1)$ theory   
with a Dirac fermion, yielding   
a $1+3$ effective theory.     
We address subtleties in the lattice fermionic action,     
such as fermion doubling and the Wilson term.     
We compute the Coleman-Weinberg potential    
for the Wilson line which is finite     
for $N$-branes $\geq 3$, due to the $Z_N$ symmetry, 
which replaces translations in the $5$th dimension.      
This mode becomes a PNGB in the low energy $1+3$ theory.   
We derive its anomalous coupling to the ``photon'' 
and its KK-modes.          
 }             
\end{abstract}              
              
\vfill              
\end{titlepage}              
              
\baselineskip=18pt              
\renewcommand{\arraystretch}{1.5}            
\pagestyle{plain}              
\setcounter{page}{1}            
%%%%%%%%%%%%%%%%%%%%%%%%%%%%%%%%%%%%%%%%%%%%%%%%%%%%%%%%%%%%%%%              
%%%%%%%%%%%%%%%%%%%%%%%%%%%%%%%%%%%%%%%%%%%%%%%%%%%%%%%%%%%%%%%%%%%%%%%%%%%%%%              
%%%%%%%%%%%%%%%%%%%%% Section 1              
%%%%%%%%%%%%%%%%%%%%%%%%%%%%%%%%%%%%%%%%%%%%              
%%%%%%%%%%%%%%%%%%%%%%%%%%%%%%%%%%%%%%%%%%%%%%%%%%%%%%%%%%%%%%%%%%%%%%%%%%%%%%              
            
\section{Introduction}

We consider a QED-like theory in $1+4$   
dimensions, periodically compactified to $1+3$. The ``electron''   
will be considered to be a heavy vectorlike   
fermion with an arbitrary mass, possibly larger   
than the compactification scale.    
We are particularly interested in the fate of the Wilson line     
in the low energy $1+4$ theory, or equivalently, the    
zero-mode of the fifth   
component of the vector potential, $A_4$ (our $1+4$ space-time indices   
run from $0$ to $4$).       
The Wilson line appears     
in the low energy $1+3$ effective     
theory as a dynamical degree of     
freedom  which imitates a low mass pseudo-Nambu-Goldstone boson     
(PNGB), and we will refer to it as the WLPNGB.    
   
In a nonabelian   
gauge theory WLPNGB's acquire (mass)$^2$ of   
order $\tilde{g}^2/R^2$ where $R$ is the size of 
the compact extra dimension and $\tilde{g}$ 
the low energy effective coupling constant,  
arising from gauge interactions   
as well as matter interactions.    
However, in a $U(1)$ gauge theory the    
WLPNGB is neutral and   
receives no contribution to its mass from gauge   
interactions. It does however, acquire mass from its coupling   
to the vectorlike fermion.    
We derive in detail the  
Coleman-Weinberg effective potential \cite{cole} of the WLPNGB,    
after dealing with a number of technical issues. 
 
We  use a lattice approximation of  
the extra dimensional theory \cite{HPW,wang0}  
(see also \cite{ACG}). 
When we latticize with $N$ slices,  
or branes, the Coleman-Weinberg potential     
for the WLPNGB is finite for $N\geq 3$. The finite  
Coleman-Weinberg potential can, moreover, be reexpressed 
in terms of the low energy parameters and is therefore 
unambiguously determined in the lattice regulator scheme. 
The minimum of the Coleman-Weinberg potential determines 
the value of the Wilson line that wraps around the 
extra dimension. This Wilson line can be absorbed 
into the the fermion field and dictates its boundary 
conditions. Under this redefinition 
we find that the minimum of the potential corresponds 
to the fermion having {\em antiperiodic boundary conditions} 
in traversing the extra dimension. 
   
The finiteness  
of the potentials for pseudo-Nambu-Goldstone     
bosons in models with $Z_N$ symmetry was noticed long ago     
\cite{georgi,ross}. 
The finiteness is a consequence     
of $Z_N$ invariance of the full theory. The     
``schizon'' models   
of Hill and Ross \cite{ross} exploited $Z_{2L}\times Z_{2R}$ 
to reduce the degree of divergence from quadratic to 
logarithmic and implement ultra-low-mass 
PNGB's to provide natural ``$5$th'' forces 
in the Standard Model, and remedy certain limits 
on the axion.  
With $Z_3$ symmetry, 
finite neutrino-schizon models have been used   
to engineer the first ``quintessence'' 
models, late-time cosmological 
phase transitions, and place limits upon time dependent 
fundamental constants \cite{hillc}.  The  
finite temperature behavior 
of such models is quite remarkable \cite{kolb}. 
Remarkably, these models are structurally equivalent     
to the present extra-dimensional scheme with 
latticized fermions when written in the   
momentum space expansion in the fifth dimension! 
 
In part, it is our interest in 
studying low energy PNGB's, such as the axion and 
ultra-low-mass cosmological PNGB's 
that has motivated the present study. 
The application of constructing models of 
ultra-low-mass PNGB's, such as the axion or quintessence, 
that are {\em immune from the effects of Planck-scale 
breaking of global symmetries,} will be described in 
a companion paper \cite{adampngb}.

In contemplating extra dimensions     
of spacetime the lattice provides a useful tool     
for regulating the enhanced quantum loop divergences of the     
extra dimensional theory, and generates a gauge     
invariant low     
energy effective Lagrangian with a finite subset      
of Kaluza-Klein modes \cite{HPW,wang0}.  A lattice description, or    
``deconstruction,''     
of a $1+4$ theory involves discretization of the     
fifth dimension, $x^4$. It therefore becomes     
an equivalent effective  $1+3$ theory with $N$   
copies of the gauge group and matter fields. 
This appears to be the 
only gauge invariant regulator for a fixed number 
$N$ of KK-modes, where $N$ plays the role of a cut-off.       
This procedure is powerful, and has suggested      
new directions and dynamics in building models     
of new physics beyond the Standard Model,     
e.g. \cite{dim,dob,susy}.   

Nevertheless, a faithful representation     
of a higher dimensional theory using a lattice involves       
subtleties which arise particularly when fermions are     
introduced. These issues can be side-stepped   
if one is only interested in a   
generalized concept of an extra dimension, e.g.,   
``theory space.'' However, we desire a {\em bona-fide} lattice   
description of a physical extra dimensional theory   
in which the lattice spectrum agrees precisely with  
the continuum spectrum at low energies, 
i.e., for $n$ KK-modes, with $n \ll N$.    
   
One of the key issues in latticizing fermions is the      
fermion flavor doubling problem. This is remedied by adding the Wilson term.     
There are also odd-even artifacts which one must reconcile.   
Finally, a redefinition of     
the parameters appearing in the     
lattice Lagrangian is required to match the continuum.     
As we will see, much of this in the $1+4$ $\rightarrow$ $ 1+3$ case     
can be understood diagrammatically.  It is in the     
fermion structure of the theory that latticization     
of an extra dimension becomes most interesting \cite{HPW,wang0}.

In addition to the Coleman-Weinberg potential, 
we also study the anomalous 
coupling of the WLPNGB to $F\;{}^\star{F}$.   
Only the WLPNGB has anomalous couplings 
to the gauge fields. There occurs a universal  
anomalous coupling of the WLPNGB to each $F_n\;{}^\star{F}_n$ 
for each KK-mode photon.

\section{QED in $5$-D Compactified to $4$-D}       
     
\subsection{Wilson Line and Gauge Invariance Under Compactification}

Consider a field which lives       
in five spacetime dimensions, $\psi(x^\mu, x^4)$,        
and which transforms        
under a local $U(1)$ gauge transformation       
$e^{i\phi(x^A)}\psi(x^A)$.  If we demand that $\psi(x^A)$ lives       
on a compact periodic manifold in the fifth       
dimension, e.g., we impose a periodic condition       
$\psi(x^\mu, x^4) = \psi(x^\mu, x^4+R)$, then we must also require       
$e^{i\phi(x^\mu, x^4)}\psi(x^\mu, x^4) =        
 e^{i\phi(x^\mu, x^4+R)}\psi(x^\mu, x^4+R)$.  This requires       
a modular constraint on the gauge transformation:       
\beq       
\phi(x^\mu, x^4+R) = \phi(x^\mu, x^4)+2\pi n.       
\eeq        
The vector potential       
$A_A$ transforms under the gauge transformation as:       
$A'_A = A_A - \partial_A\phi$.       
Consider the Wilson line around the periodic manifold:       
\beq       
\chi(x^\mu) = \int_0^R dx^4 A_4(x^\mu, x^4).       
\eeq      
The Wilson line, $\chi(x^\mu)$ behaves like a  
dynamical spin-$0$ field in the $1+3$ theory. Indeed, 
$\chi$, given a canonical normalization,  
is the Wilson-line pseudo-Nambu-Goldstone boson (WLPNGB).   
Under the gauge transformation the Wilson line transforms as:       
\beq       
\chi(x^\mu) \rightarrow \int_0^R dx^4 A'_4 = \chi(x^\mu) - 2\pi n.    
\eeq     
The Wilson line is just the ($p^4=0$) zero-mode of $A_4$. 
Under gauge transformations the zero-mode can only 
be shifted by a constant.     
Owing to the periodic compactification, gauge transformations   
of the zero mode are thus restricted:   
\beq       
 A'_4(x^\mu) = A_4(x^\mu) - 2\pi n/R.    
\eeq  
These results imply that the  
continuous $U(1)$ ``chiral symmetry'' 
of the $\chi$ field is explicitly broken 
to the center $Z_N$, and 
in general $\chi$ will acquire a mass.   
A lattice regularization of the quantum loops 
of the theory 
manifestly respects this constraint.   
 
As we will see, with the lattice regulator, the $Z_N$ symmetry,   
takes the place of the continuous $S_1$ translational  
symmetry of the continuum limit. The Wilson line 
effective potential is most naturally computed 
in a lattice approximation.

\subsection{Gauge Field Lattice}      
      
Presently we consider a $U(1)$ gauge theory     
in $1+4$ dimensions that is {\em periodically     
compactified} to $1+3$.       
 The latticization  of a $U(1)$     
gauge theory with a periodic fifth dimension is straightforward.          
The effective Lagrangian becomes      
the gauged chiral Lagrangian in $1+3$ dimensions for $N$ copies       
of the $U(1)$ gauge group:        
\be      
\label{lattone}       
{\cal{L}}= \sum_{n=1}^N\; \left[- \frac{1}{4} F_{n\mu\nu} F_n^{\mu\nu}        
   + D_{\mu}\Phi_n^\dagger D^{\mu}\Phi_n -V(\Phi_n)\right].        
\ee        
Here we have $N$ gauge groups $U(1)_n$ with      
a common (dimensionless) gauge coupling  $g$,       
and $N$ link-Higgs fields,       
$\Phi_n$, having charges $(0,0,...,-1_n,1_{n+1},...,0)$  
under $(U(1)_1,...,U(1)_N)$.      
The covariant derivative acts     
upon $\Phi_n$ as:     
\beq       
D_{\mu}\Phi_n = \partial_{\mu}\Phi_n +   
i g(\tilde{A}_{{n}\mu}-\tilde{A}_{{n+1}\mu}) \Phi_n.     
\eeq   
where all fields are functions of $1+3$ spacetime  
$x^\mu$.  
(Note: Henceforth  
we will denote the $x^4$ configuration space vector potentials  
$\tilde{A}_{{n}\mu}$ and Higgs-phase fields $\tilde{\chi}_n$ with a  
tilde; the corresponding fields without the tilde will be  
conjugate $p^4$ momentum space fields, ${A}_{{p}\mu}$  
and $\chi_p$; $n=N+1$ is identified with $n=1$.) 
  
As a boundary condition, we identify:      
\beq   
\label{bc}   
\Phi_{n}= \Phi_{mN+n}     
\eeq      
for integer $m$.  This implements the periodic compactification. 
Notice that eq.~(\ref{lattone}) with eq.~(\ref{bc}) is $Z_N$ invariant under 
$F_{n\mu\nu}\rightarrow F_{(n+m)\mu\nu}$ 
and $\Phi_n\rightarrow \Phi_{(n+m)}$.  This $Z_N$ invariance 
has replaced the continuous $S_1$ translational invariance 
of the compactified theory. It is an arbitrarily good approximation  
for large $N$ in physical quantities that are insensitive to  
the short-distance (UV) structure of the theory. The Coleman-Weinberg 
potential is such a UV-safe quantity. 
 
The potential $V(\Phi_n)$ causes each $\Phi_n$ 
to develop a common VEV.        
$\Phi_n$ is then interpreted as the Wilson     
link, linking brane $n$ to brane $n+1$:     
\beq     
\label{wilson}     
\Phi_n = (v/\sqrt2g)\exp\left(ig\int_{x^4_n}^{x^4_{n+1}} dx^4 \tilde{A}_4\right)     
\rightarrow (v/\sqrt2g)\exp(iga \tilde{A}_{4n}).     
\eeq  
Thus, $\Phi_N$ links brane $N$ to brane $1$.    
Here $a$ is     
the physical lattice constant, the distance between     
nearest neighboring branes in $x^4$.  In order    
for eq.~(\ref{lattone}) to reproduce     
the continuum limit kinetic terms,      
$-(1/4)(\partial_\mu \tilde{A}_4 -\partial_4 \tilde{A}_\mu)^2$,     
we must take     
$v=1/a$, which is related to the compactification scale:     
\beq     
\label{compact}     
R = Na = N/v.     
\eeq

From the point of view of $1+3$ dimensions, each $\Phi_n$       
is effectively a nonlinear-$\sigma$ model field:        
\be      
\label{phi}       
\Phi_n \rightarrow (v/\sqrt{2}g)\exp(ig\tilde{\chi}_n / v).         
\ee       
The $\Phi_n$ kinetic terms then go over to a mass-matrix     
for the gauge fields:     
\be     
\label{massmatrix}     
\half v^2\sum_{n=1}^N  \left( (\tilde{A}_{n+1\mu}- \tilde{A}_{n\mu}) -     
 \frac{1}{v}\partial_\mu\tilde{\chi}_n      
\right)^2.     
\ee      
To diagonalize eq.~(\ref{massmatrix}) it is useful to pass      
to a complex representation.     
Without loss of generality consider:     
\bea     
\tilde{A}_{n\mu}& = &     
  \frac{1}{\sqrt{N}}\sum_{p=-J}^{J+\delta}\; A_{p\mu} \exp(2\pi i p n/N);     
\qquad A_{p,\mu}^* = A_{-p,\mu},    
\\     
\tilde{\chi}_n & = &   
\frac{1}{\sqrt{N}}\sum_{p=-J}^{J+\delta} \; \chi_p \exp(2\pi i p n/N);     
\qquad \chi_p^* = \chi_{-p}.     
\eea     
Here $J=(N-1)/2$ and $\delta=0$ for $N$ odd      
[$J=(N-2)/2$ and $\delta = 1$ for $N$ even]. 
The $p$-representation preserves canonical 
normalizations, for example  
the $U(1)$ kinetic terms in the Lorentz 
gauge, $\partial_\mu A^\mu_n=0$, become: 
\bea     
-\frac{1}{4}\sum_{n=1}^N \;F_{n\mu\nu}F^{\mu\nu}_n 
& = &-\half(\partial_\mu A_{0\nu})^2   
-\sum_{p=1}^{J} |\partial_\mu A_{p\nu}|^2  
-\left\{ \half {\delta} \right\}(\partial_\mu A_{(N/2)\nu})^2 .    
\eea  
Note the last term is absent when $N$ is odd ($\delta=0$). 
Henceforth, for notational 
simplicity, we will refrain from writing the $\delta$ 
term, though it is implicitly present when $N$ is even.  
Now, we define:     
\beq     
F_p = [\exp(2\pi i p/N)-1],\qquad F_p^*F_p = 4\sin^2(\pi p/N) ,    
\eeq     
and then see that eq.~(\ref{massmatrix}) becomes:     
\bea     
& &      
\half (\partial_\mu\chi_0)^2 + \sum_{p=1}^{J} \;      
v^2|A_{p\mu}F_p - \frac{1}{v} \partial_\mu\chi_p |^2 .     
\eea     
We can perform a momentum space     
gauge transformation on each component,     
except $p=0$:     
\beq     
{A}_p \rightarrow A_p + \frac{1}{v F_p } \partial_\mu\chi_p\qquad p\neq 0;     
\qquad {A}_0 \rightarrow A_0.     
\eeq     
From this we obtain:     
\beq     
\half(\partial_\mu\chi_0)^2 + 
 4v^2\sum_{p=1}^{J}|{A}_{p\mu}|^2 \sin^2(\pi p/N) . 
\eeq     
The spectrum therefore contains the real zero-mode of $A_\mu$, 
$A_{4}$ (which is  $\chi_0$), and a tower of     
doubled Kaluza-Klein-modes, appearing as massive photons,    
each pair labeled by $p$, of mass:     
\beq     
M_p^2 = 4v^2 \sin^2(\pi p /N).     
\eeq   
For $N$ even ($\delta=1$)  
the $p=N$ mode occurs as well, as a singlet 
with (mass)$^2$ $4v^2$.   
Correspondingly, all but the zero-mode     
linear combination of the $\chi_0$ are ``eaten''     
to become longitudinal modes.

\begin{figure}[t]     
\vspace{7cm}     
\includegraphics{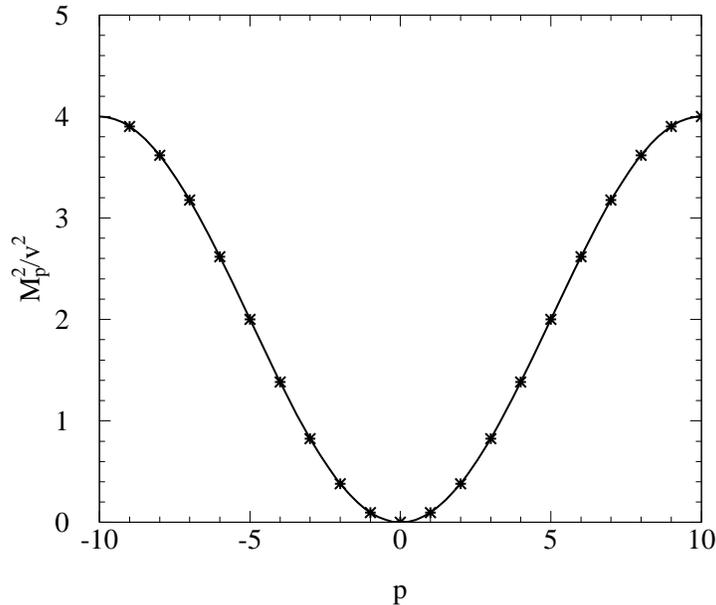}     
\vspace{1cm}     
\caption[]{\small \addtolength{\baselineskip}{-.4\baselineskip}     
The vector boson mass spectrum is plotted for $N=20$ and $N=19$.    
For $N=19$ all states are as indicated from $-9\leq p\leq 9$, including   
the single $p=0$ zero-mode photon.    
For $N=20$ we include an extra state   
$p=10$.  The spectrum exhibits    
physical doubling, a consequence of periodic compactification, and   
encompasses a single Brillouin zone.    
 }     
\label{vmass}     
\end{figure}

The     
spectrum of gauge fields for periodic     
compactification was discussed previously in \cite{wang0}.     
For $N$ odd [even] it admits a zero mode gauge field, and     
a tower of $(N-1)/2$ [$(N-2)/2$] doubled Kaluza-Klein modes [and     
a singlet highest mode]. This doubling is normal    
and a physical consequence of the periodic compactification;    
e.g., there will occur both left moving and right moving    
modes in the periodic manifold and these are degenerate    
(alternatively, the sine modes are degenerate with    
the corresponding cosine modes). This is shown in   
Figure \ref{vmass} for $N=20$ and $N=19$. The      
first Kaluza-Klein gauge mode has a mass of     
$M_K \approx 2v\pi /N$ \cite{wang0}. This is identified     
with the compactification scale $2\pi/R$, hence we again     
recover eq.~(\ref{compact}), $vR=N$.     
 
In summary, 
the master gauge group,       
$U(1)^N$, is broken to the diagonal subgroup $U(1)$     
by the $\Phi_n$.  $N-1$ of the       
link fields $\chi_n$ are eaten by the Higgs mechanism, giving $N-1$       
massive vector fields (the Kaluza-Klein states).  We       
are left with one massless vector field and one massless scalar.  The       
linear combination of the link fields that remains massless in     
the classical limit is:       
\begin{equation}       
\chi_0 \equiv -\frac{iv}{g\sqrt{N}}      
  \ln \left[ \Pi_{n=1}^N(\sqrt2g\Phi_n/v) \right] =       
\frac{1}{\sqrt{N}}\sum_{n=1}^N \tilde{\chi}_n.       
\end{equation}       
This mode is the WLPNGB.

\subsection{Latticizing Fermions}     
     
We now include a single Dirac fermion in     
the $1+4$ theory with charge $q=-1$.     
The fermion-gauge boson system has     
the continuum  action:     
\beq     
\label{contferm}     
{\cal{L}} = \int d^5x \; \bar{\Psi}[(i\slash{\partial}-g\slash{A})     
-(\partial_4+igA_4)\gamma^5 - M ]\Psi.     
\eeq    
where the fifth $\gamma$-matrix is $i\gamma^5$.  
Consider $A_4=0$.     
For the fermion in the continuum with periodic     
compactification of the $x^4$ spatial manifold      
we can impose a periodic boundary conditions:     
\beq     
\Psi(x^4) = \psi(x^4+R).     
\eeq     
If, however,  we view the manifold as the     
{\em boundary} of a $1+5$ dimensional space then     
we must use the {\em antiperiodic} boundary condition:     
\beq     
\Psi(x^4) = -\psi(x^4+R).     
\eeq     
In the latter case the minus sign arises because     
the spinor is being rotated through $2\pi$ in the $1+5$     
space as we traverse the periodic compactification.   
 
In the periodic case the      
$x^4$ momentum space basis     
functions are $\exp(ikx^4)$ where the momenta      
are $k = (2p)\pi /R$ where $p $ is     
an integer that runs from     
$-\infty$ to $+\infty$. The $p=0$ fermion     
is potentially a zero mode, but will always   
be vector-like (having both $L$-- and $R$--   
handed components).     
In the antiperiodic case, the $x^4$  momenta      
are $k = (2p+1)\pi /R$ where $p $ is     
an integer that runs from     
$-\infty$ to $+\infty$.      
There is no zero mode in the antiperiodic case.     
   
In both periodic and antiperiodic boundary     
conditions all     
levels other than the zero mode     
are doubled as $(p_1,p_2)$: (periodic) $(-1,1)$ $(-2,2)$;     
(antiperiodic) $(-1,0)$, $(-2,1)$,     
etc.       
This doubling is physical,    
as we saw in the gauge field case,    
corresponding to the fact that a traveling     
wave packet moving to the left is degenerate     
with one moving to the right. The physical    
doubling  stems from parity, i.e.,     
the symmetry of $x^4\rightarrow -x^4$.

Consider now a nontrivial Wilson line,     
\beq     
g\int_0^R dx^4\; A_4 = \pi; \qquad       
A_4 = \frac{\pi}{gR},     
\eeq     
which can be implemented with the     
indicated constant gauge field.     
If we begin with the antiperiodic boundary condition,     
we can nonetheless redefine the fermion field     
by a phase factor as:     
\beq     
\hat{\Psi}(x^4) = \exp \left( ig\int_0^{x^4} dx^4\; A_4 \right)\Psi(x^4).     
\eeq     
With this redefinition we thus ``gauge away'' the Wilson line, but     
we have now modified the fermion boundary condition     
to be periodic.       
Indeed, we are free to use the Wilson line     
to achieve any desired boundary conditions     
on the fermion field.  The presence of the $U(1)$     
interaction thus influences the spectrum of     
fermions in a fundamental way.  As we will see, however,     
the ground-state energy, when minimized, will determine     
the particular value of the Wilson line, hence the     
fermionic boundary conditions, dynamically.      
(Of course, if we view the $1+4$ manifold as a boundary of     
$1+5$, then there is additional field energy     
contributing through the magnetic field     
living in the $1+5$ bulk.)  We will find that the {\em antiperiodic 
boundary conditions} are dynamically favored.   
     
The fermionic $U(1)$ theory     
can be latticized as follows.  We place 
independent fermions, $\psi_n$ on brane $n$, 
i.e., having charge $(0,0,..., -1_n,...., 0,0 ..)$, hence  
the covariant derivative acts as: 
\beq 
D_\mu \tilde\Psi_n =(\partial_\mu + ig \tilde{A}_{n\mu})\tilde\Psi_n . 
\eeq  
Then a ``naive'' lattice in $x^4$ leads to 
the action:    
\beq     
\sum_{n=1}^N\int d^4x \;\left[ \bar{\tilde\Psi}_n(i\slash{D}     
 - M )\tilde\Psi_n -      
\left(\frac{1}{\sqrt{2}} g\eta  \bar{\tilde\Psi}_n \gamma_5 \Phi_{n}  
\tilde\Psi_{n+1} + h.c. 
\right) \right]    
 \label{latferm}     
\eeq     
The nearest neighbor hopping term represents   
the kinetic term in the fifth dimension.   
 We have taken the     
simplest lattice approximation to the derivative,     
\beq     
\partial_4\Psi \rightarrow (\tilde\Psi_{n+1} -\tilde\Psi_n)/a .     
\label{der}    
\eeq     
Note that the normalization 
of the hopping term follows from the normalization of   
$\Phi_n = v/g\sqrt{2} $, when $A_4=\chi_n=0$,    
together with $v=1/a$, and 
the fact that we latticized a Hermitian ``backward-forward''    
derivative, $\partial_4 \equiv (-i/2)(\stackrel{\rightarrow}{\partial}_4 
-\stackrel{\leftarrow}{\partial}_4)$,  
which together with the $i\gamma^5$ leads    
to a Hermitian kinetic term (or equivalently, use    
the ``forward'' derivative $(i/2)\stackrel{\rightarrow}{\partial}$ 
of  eq.~(\ref{der})    
and add $+h.c.$); this kills off the     
diagonal $\bar\psi_n\gamma^5\psi_n$ terms.    
Here we have introduced a parameter $\eta$     
which will be needed for matching to the  
continuum theory spectrum.  
   
\begin{figure}[t]     
\vspace{7cm}     
\includegraphics{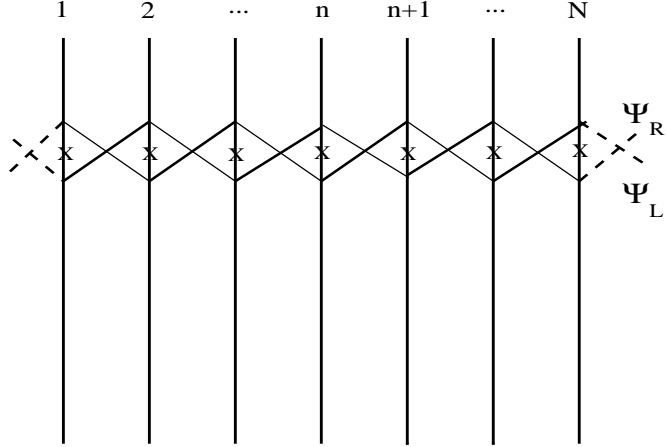}     
\vspace{1cm}     
\caption[]{\small \addtolength{\baselineskip}{-.4\baselineskip}     
Dirac fermion      
has both chiral modes on each brane, where the upper (lower) 
vertices are $R$ ($L$) modes. 
The $\times$'s denote the     
$M$ terms on each brane which couple 
$\tilde\Psi_{nR}$ and $\tilde\Psi_{nL}$, and the cross-bars are the      
latticized fermion kinetic (hopping) terms,  
$\bar{\Psi}_n\Phi_n{\tilde\Psi}_{n+1}$     
couplings.      
The spectrum has a singlet lowest massive     
mode of mass $M$, and doubled Kaluza-Klein modes; by adding a      
Wilson term one can remove one of the     
two cross-bars between adjacent branes, and eliminate      
fermion doubling     
in the spectrum \cite{wang0}.}     
\label{dirac1}     
\end{figure}

Passing to a chiral projection basis     
on each brane,  eq.~(\ref{latferm}) can be written     
as:     
\bea     
& & \sum_{n=1}^N\int d^4x \;\left[  \bar{\tilde\Psi}_{nL}(i\slash{D}     
 )\tilde\Psi_{nL}   
 +    \bar{\tilde\Psi}_{nR}(i\slash{D}     
)\tilde\Psi_{nR} - M (\bar{\tilde\Psi}_{nL}\tilde\Psi_{nR} + h.c.)  \right] \nonumber \\   
 & &    
-\frac{1}{\sqrt{2}}  \sum_{n=1}^N\int d^4x \; g\eta  
\left[\bar{\tilde\Psi}_{nL} \Phi_n\tilde\Psi_{(n+1)R} -     
\bar{\tilde\Psi}_{nR} \Phi_n \tilde\Psi_{(n+1)L} + h.c.\right]     
 \label{latferm1}     
\eea     
This is illustrated in Figure \ref{dirac1}.  We obtain     
the ``zig-zag'' pattern as each chirality      
hops to the opposite chirality     
on the neighboring brane.

Let us, for the sake of discussion, impose antiperiodic     
boundary conditions on the fermion field:     
\beq     
\label{odd}     
\tilde\Psi_{n} \equiv (-1)^m\tilde\Psi_{n + mN}     
\eeq      
for integers $n,m$  (Note that the action remains 
of $Z_N$ invariant).  
The mass spectrum of eq.~(\ref{latferm1}) is     
derived by diagonalizing the Lagrangian. Let:     
\beq     
\label{expand}   
\tilde\Psi_n = \frac{1}{\sqrt{N}}\sum_{p=-J}^{J+\delta} e^{(2p+1)i\pi n/N} \Psi_p.     
\eeq     
%where $J= (N-1)/2$ for $N$ odd      
%and $J=N/2$ for $N$ even.      

\begin{figure}[t]     
\vspace{7cm}     
\includegraphics{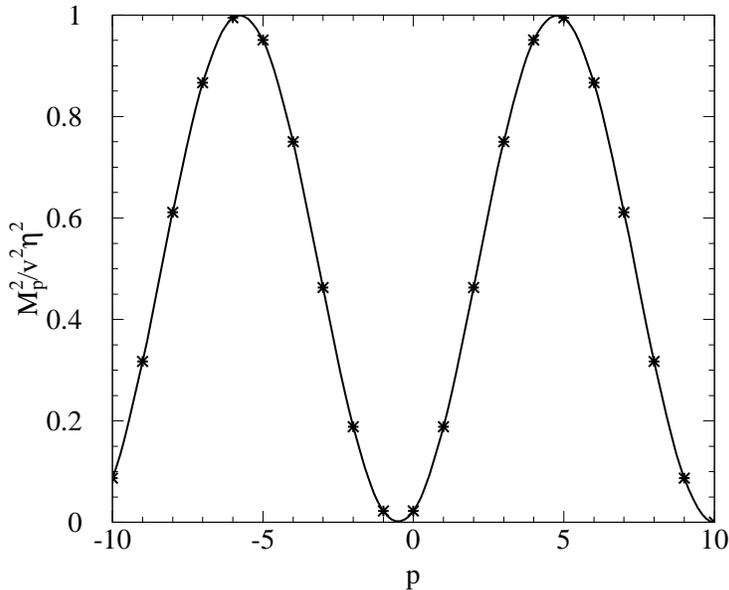}     
\vspace{1cm}     
\caption[]{\small \addtolength{\baselineskip}{-.4\baselineskip}    
The fermion mass spectrum, with $M=0$ and without   
the Wilson term (in   
units of $v^2 \eta^2$).  Here we choose $N=21$ odd.    
The spectrum exhibits    
physical doubling within each minimum,    
a consequence of periodic compactification. It   
encompasses two Brillouin zones.  The state   
with $p=10$ (which is equivalent to $p=-10$)    
has the lowest mass, but   
represents an unphysical second flavor, the usual fermion doubling problem   
for latticized fermions.    
 }     
\label{fmass2}     
\end{figure}

\begin{figure}[t]     
\vspace{7cm}     
\includegraphics{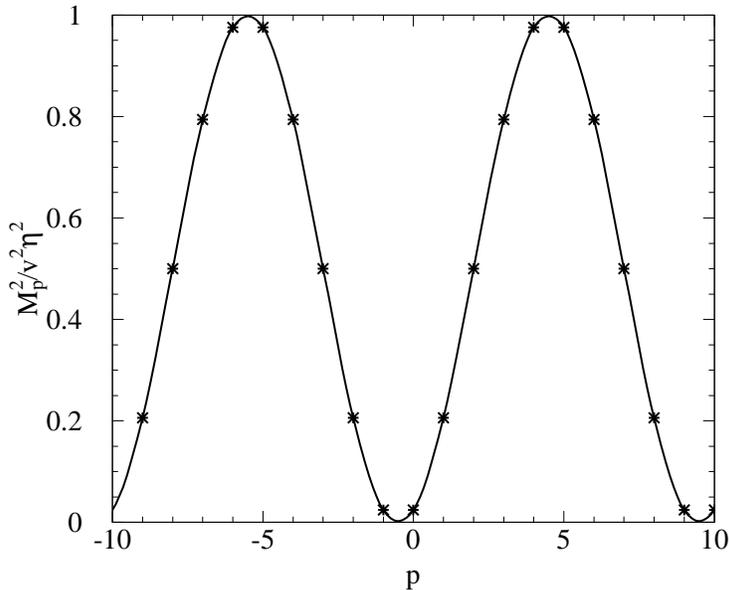}     
\vspace{1cm}     
\caption[]{\small \addtolength{\baselineskip}{-.4\baselineskip}    
The fermion mass spectrum, with $M=0$ and without   
the Wilson term (in   
units of $v^2 \eta^2$).  Here we choose $N=20$ even.    
Notice the quadrupling of levels and the absence   
of a fermionic zero mode due to the   
antiperiodic boundary conditions.    
The spectrum exhibits    
physical doubling within each minimum,    
a consequence of periodic compactification. It   
encompasses two Brillouin zones, however, leading to an overall    
quadrupling of the spectrum.  The states in the vicinity   
of $p=10$ are an unphysical second flavor.    
 }     
\label{fmass}     
\end{figure}

We see that eq.~(\ref{odd})     
is therefore implemented. Note that there is no     
value of $p$ for which $2p+1$ vanishes, hence     
there is no fermionic zero mode with    
antiperiodic boundary conditions, as in the continuum case. 
Eq.~(\ref{latferm}) becomes:     
\beq     
\sum_{p=-J}^{J+\delta}\int d^4x  
\;\left\{ \bar{\Psi}_p(i\slash{\partial} - M )\Psi_p  
- i(\eta v) \sin\left[{(2p+1)\pi/N}\right] 
\bar{\Psi}_p \gamma_5 \Psi_{p} \right\} ,     
 \label{latferm2}     
\eeq     
where we have substituted eq.~(\ref{phi}) with $\chi_n = 0$ into this     
expression, and suppressed the ${A}_{q\mu}$ in the covariant 
derivatives.     
The     
mass of the $p$-th mode is therefore:     
\beq     
M_p^2 = M^2 + \eta^2 v^2\sin^2\left[(2p+1)\pi/N\right].     
\eeq     
Naively, we would argue     
that the low lying levels of this spectrum     
expanded about $p=0$ match onto the      
continuum with the matching condition of     
eq.~(\ref{compact}).    
Owing to the fermionic antiperiodic boundary condition   
the lowest modes in the continuum   
have  masses $M^2_p =  M^2 + (2p+1)^2\pi^2/R^2$.   
We obtain presently, for the lattice theory with   
small $p$, $M^2_0 =  M^2 + \eta^2v^2(2p+1)^2\pi^2/N^2$.   
Thus, using $v = 1/a =N/R$, the matching of the low energy modes     
to the continuum requires a      
choice of $\eta=1$.    
However, we must     
examine the spectrum of the lattice theory     
in greater detail.

First, consider $N$ odd (see Fig.~(\ref{fmass2})), and  
for simplicity we   
presently take $M=0$ and $\eta=1$.     
Then we see that the lowest mass     
fermion is the $p=(N-1)/2$ mode,     
which is the lowest mass     
state of zero mass, and is      
an undoubled singlet.     
This state was absent in the continuum case     
and must be a lattice artifact.     
All other modes are doubled.     
The $p=0$ mode is the next     
lightest state and is degenerate with $p=-1$     
with mass $v^2\sin^2(\pi/N)$;     
the $p=-(N-1)/2$ with $p=(N-3)/2$     
with mass     
$v^2\sin^2(2\pi/N)$, etc.       
This doubling     
is just the conventional $x^4\rightarrow -x^4$     
invariance.     
 The mass as a function      
of $p$, therefore, has two basins about the two distinct     
minima of $\sin^2((2p+1)\pi/N)$      
corresponding to the      
towers of states: $p=[(N-1)/2][-(N-1)/2,(N-3)/2]...$ and     
$p=[0,-1],[1,-2] ...$     
These two basins of states correspond to     
two distinct Brillouin zones. This is the      
familiar lattice fermion flavor doubling problem.     
     
We {\em cannot} interpret the $p=(N-1)/2$ mode     
as the ground state of a tower with the      
$p=[0,-1]$ modes     
as next in the sequence. The transition     
from $p=0$ state of mass $\approx v\pi/N $     
to the $p=(N-1)/2$     
of mass $=0$     
is allowed virtually with the emission     
of a $p=[(N-1)/2]$ heavy photon.      
However the mass of this photon is     
$2v\sin(\pi (N-1)/2N) \approx 2v$,     
and the transition can never match energetically,     
even approximately.       
On the other hand, the transition     
from the $p=1$ state of mass $\approx 3v \pi/N$     
to $p=0$ of mass $\approx v\pi/N $     
does match the $p=1$ photon     
of mass  $2v\sin(\pi/N) \approx 2v\pi/N$.     
Of course,    
these energetics only make sense in the $N\rightarrow \infty $     
limit.  Hence, we must conclude that the second 
basin of states (Brillouin zone) represents 
a second spurious fermionic flavor.

For $N$ even (see Fig.~(\ref{fmass})),     
we see that the      
ground-state fermion mode $p=0$ is      
a non-zero mode, and there is a     
4-fold degeneracy with     
the $p=-1$, $p=N/2$, $p=N/2-1$ modes;     
the $p=1$ is degenerate with $p=-2$ and  the     
$p=N/2+1$, $p=N/2-2$ modes, etc,     
spanning the tower of states sequentially.      
Again, these form two distinct     
basins,     
$p=[0,-1],[1,-2],...$ and      
$p=[N/2, N/2-1], [-N/2+1, N/2-2],...$     
with accidental degeneracy, and represent two  
distinct flavors.     
     
The fermion flavor     
doubling behavior can be understood graphically 
from Figure \ref{dirac1}.     
Consider the zig-zag hopping line emanating from $\Psi_{R1}$,     
coursing through $\Psi_{L2},\Psi_{R3}, ... , \Psi_{LN}$.     
The accidental degeneracy     
for $N$ even arises because     
the line which started on $\Psi_{1R}$     
ends on $\tilde\Psi_{nL}$.  When we make the periodic hop     
we have $\Psi_{LN}$ rejoining to the starting point,     
$\Psi_{R1}$. Similarly,     
$\Psi_{L1}$ is rejoined by its starting point     
$\Psi_{RN}$. Hence there are     
two independent zig-zag hopping paths that course through the lattice. 
For $M=0$ these are two independent flavors, coupled only 
by chirality conserving gauge field interactions.     
These are equivalent under the $Z_2$ transformation (parity)     
which swaps $L$ and $R$, hence there is quadrupling.     
With $N$ odd, there is only one zig-zag line     
since, starting from $\Psi_{R1}$, one     
reaches $\Psi_{RN}$ and rejoins $\Psi_{L1}$, etc.     
Thus, the quadrupled degeneracy is lifted, and the ground state     
is a singlet, and all other modes are doubled.     
This lifting of degeneracy is a small effect     
in the large $N$ limit. In this limit      
we just ignore the connection from $\Psi_N$ to $\Psi_1$     
then the two zig-zag paths are degenerate. This     
then leads to the fermion flavor doubling in the spectrum.     
     
As mentioned above, the usual lattice fermion flavor     
doubling is always present with the simplest hopping,     
or first order approximation to the derivative, and     
the second Brillouin zone      
then appears  in the spectrum. The     
Brillouin zones are defined by the basins in momentum     
space localized around minima of the energy (mass),     
and each tower of states built around the minima.

Notice that, with the Dirac     
mass $M\neq 0$ and large,    
these transition energetics can never match, even in     
the continuum theory!  For example, expanding in $v/M$ for   
large $M$, we can     
consider the  $p=1$ state of mass $\approx M + 9v^2 \pi^2/2N^2M$   
decaying to the     
$p=0$ state of mass $\approx M + v^2\pi^2/2N^2M $. The transition   
energy is thus $\approx 4v^2\pi^2/N^2 M$  and cannot   
match the $p=1$ photon     
of mass  $2v\sin(\pi/N) \approx 2v\pi/N$.    
We emphasize that this {\em is not a lattice   
artifact}! A heavy fermion will     
have slow, virtually mediated, decays of its KK-modes,   
i.e., all KK-modes become quasistable in the large   
$M$ limit.     
    
This discussion is {\em not to imply} 
that we cannot build a model in which the lattice is 
real, and the second 
Brillouin zone is therefore physical, e.g., perhaps it is possible 
to interpret sequential generations of flavors in this way (!)  
For the present discussion, however,  
we are interested in a faithful lattice 
representation of the continuum, hence we view 
the flavor doubling as an unwanted artifact.

\subsection{Incorporating the Wilson Term}

\begin{figure}[t]     
\vspace{5cm}     
\includegraphics{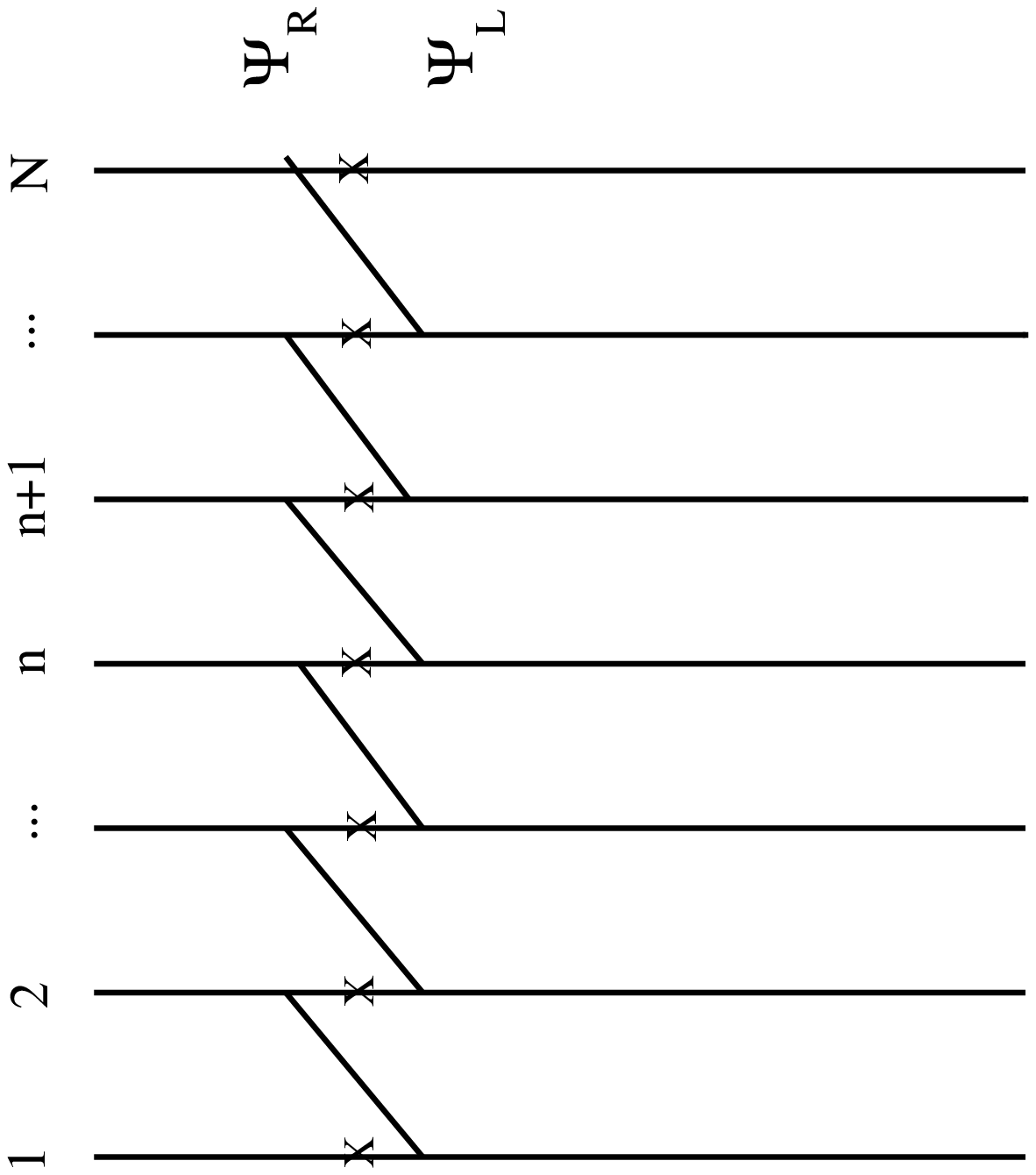}     
\vspace{6.5cm}     
%\caption[]{\small \addtolength{\baselineskip}{-.4\baselineskip}     
% \cite{wang0}.}     
%\label{dirac2}     
%\end{figure}     
%\begin{figure}[t]     
%\vspace{6cm}     
\includegraphics{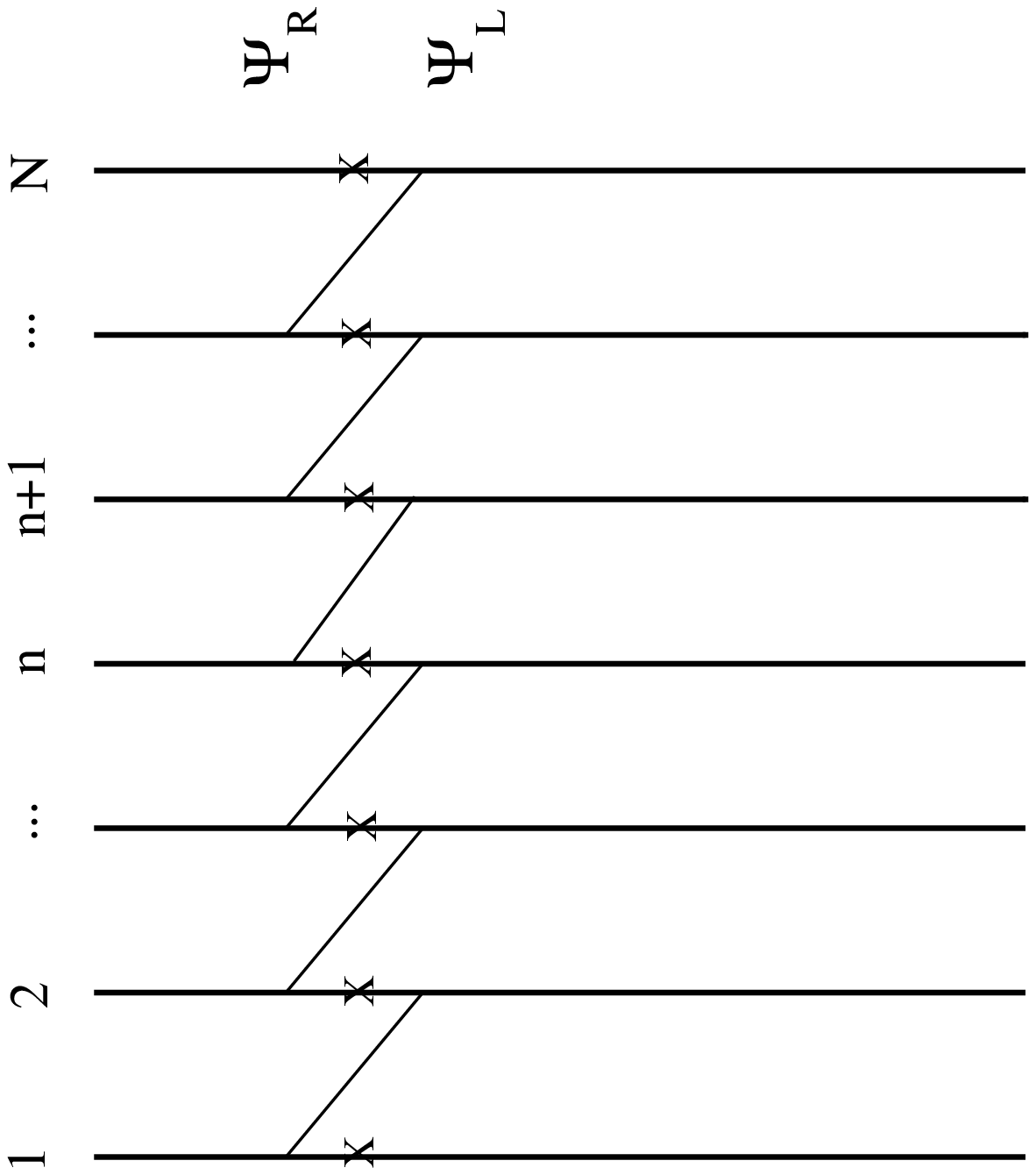}     
\vspace{0.2cm}     
\caption[]{\small \addtolength{\baselineskip}{-.4\baselineskip}     
Adding a Wilson term      
annihilates half of the links. The choice of lattice   
is controlled by the the relative sign, $\eta' = \eta$   
($\eta' = -\eta$)   
yields the top (bottom) figure.   
The doubling problem     
is now solved. Both lattices are periodic \cite{wang0}.}     
\label{dirac2}     
\end{figure}

The lattice artifact problems described above    
have a well-known     
remedy, the addition of     
the ``Wilson term''  (see, e.g.,  the lectures 
of A. Kronfeld  in \cite{kron}). 
The Wilson term is a higher dimension operator     
that we add to the continuum theory     
of eq.~(\ref{contferm}), which acts like a bosonic 
kinetic term:     
\beq     
\label{contferm2}     
\int d^5x \; \bar{\Psi}[(i\slash{\partial}-g\slash{A})     
-(\partial_4+igA_4)\gamma^5 - M - \frac{1}{M_X}(\partial_4+igA_4)^2]\Psi.     
\eeq    
The sign of the Wilson term is fixed by positivity of the   
energy, while the relative sign of the $\gamma^5$
term is arbitrary.     
In the lattice theory this amounts to adding a term to     
eq.~(\ref{latferm1}) of the form:     
\beq     
-\sum_{n=1}^N\int d^4x \;      
\frac{\eta'}{2a^2 v M_X}     
\left(\sqrt{2} g\bar{\tilde\Psi}_n \Phi_{n}\tilde\Psi_{n+1} 
+ \sqrt{2}g\bar{\tilde\Psi}_{n}\Phi_{n-1}^{\dagger}\tilde\Psi_{n-1}     
-2v\bar{\tilde\Psi}_{n}\tilde\Psi_{n}\right) .     
\label{latferm3}     
\eeq     
We have installed a coefficient $\eta'$ which will be determined   
momentarily. For convenience    
we define $M_X= 1/a\equiv v$.      
Eq.~(\ref{latferm3}) then takes the form in the chiral basis:     
\beq     
-\half \eta' \sum_{n=1}^N\int d^4x \;\left(      
\sqrt{2}g\bar{\tilde\Psi}_{nL} \Phi_n\tilde\Psi_{(n+1)R}+   
\sqrt{2}g\bar{\tilde\Psi}_{nR} \Phi_n\tilde\Psi_{(n+1)L}     
-2 v \bar{\tilde\Psi}_{nL}\tilde\Psi_{n,R} + h.c. \right)    
 \label{latferm4}     
\eeq     
Note that we conjugated and   
used the shift symmetry in the sum $n\rightarrow n+1$   
for the second term above.         
Adding the Wilson term then modifies eq.~(\ref{latferm1}):   
 \bea     
& & \sum_{n=1}^N\int d^4x \; \left[ \bar{\tilde\Psi}_{nL}(i\slash{D}     
 )\tilde\Psi_{nL}   
 +    \bar{\tilde\Psi}_{nR}(i\slash{D}     
)\tilde\Psi_{nR} - \widetilde{M} (\bar{\tilde\Psi}_{nL}\tilde\Psi_{nR} + h.c.) \right]  \nonumber \\   
 & &    
-\frac{1}{\sqrt{2}} \sum_{n=1}^N\int d^4x \;   
\left[g(\eta'+\eta)\bar{\tilde\Psi}_{nL} \Phi_n\tilde\Psi_{(n+1)R}     
 + g(\eta'-\eta)\bar{\tilde\Psi}_{nR} \Phi_n\tilde\Psi_{(n+1)L}     
+ h.c.\right]     
 \label{latfermw}     
\eea    
where:   
\beq   
\widetilde{M} = M-v\eta'  . 
\eeq   
We can thus choose $\eta'=\eta$ to cancel the $nR \rightarrow (n+1)L$  
hopping terms,   
or  $\eta'=-\eta$ to cancel the $nL\rightarrow (n+1)R$ hopping 
terms.  Note that our freedom  to choose the     
sign of the $(\partial_4+igA_4)\gamma^5$ term relative     
to the Wilson term by the $Z_2$ parity inversion of $\gamma_5$,   
amounts to a freedom of choice in the sign of $\eta$.    
This flips $L$ and $R$ in   
eq.~(\ref{latfermw}) and allows  Figure \ref{dirac1} to be replaced by either 
the upper or lower of Fig.~(\ref{dirac2}).

\begin{figure}[t]     
\vspace{7cm}     
\includegraphics{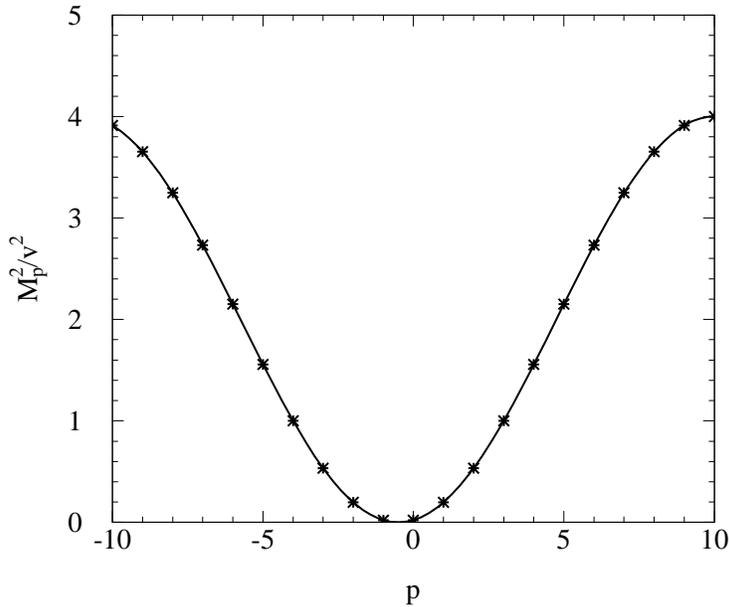}     
\vspace{1cm}     
\caption[]{\small \addtolength{\baselineskip}{-.4\baselineskip}    
The fermion mass spectrum, with $M=0$ and with   
the Wilson term (in   
units of $v^2 \eta^2$).  Here we choose $N=21$ odd.    
The spectrum exhibits    
physical doubling     
a consequence of periodic compactification. It   
encompasses now only one Brillouin zones, eliminating   
the unphysical second flavor.    
 }     
\label{wmass2}     
\end{figure}

Let us choose $\eta=\eta'$ and replace $\Phi_n = v/\sqrt{2}g$,   
whence the Lagrangian eq.~(\ref{latfermw}) becomes:   
 \bea     
& & \sum_{n=1}^N\int d^4x \; \left[ \bar{\tilde\Psi}_{nL}(i\slash{D}     
 )\tilde\Psi_{nL}   
 +    \bar{\tilde\Psi}_{nR}(i\slash{D}     
)\tilde\Psi_{nR} - \widetilde{M} (\bar{\tilde\Psi}_{nL}\tilde\Psi_{nR} + h.c.) \right]  \nonumber \\   
 & &    
-\sum_{n=1}^N\int d^4x \;     
\left[\eta v \bar{\tilde\Psi}_{nL}\tilde\Psi_{(n+1)R}     
+ h.c.\right]     
 \label{latfermw2}     
\eea       
In the momentum eigenbasis, eq.~(\ref{expand}),    
with $\Phi_n= v/g\sqrt{2}$, the mass term becomes:   
\beq     
-\sum_{p=-J}^{J+\delta}\int d^4x \; \left[\bar{\Psi}_p(     
 \widetilde{M} + \eta v \cos\left[{(2p+1)\pi/N}\right])\Psi_p +      
i\eta v \sin\left[{(2p+1)\pi/N}\right]\bar{\Psi}_p \gamma_5 \Psi_{p}\right] ,     
 \label{latferm3a}     
\eeq     
The spectrum with the Wilson term becomes:     
\beq     
\label{wilsonterm}     
M_p^2 = M^2 + 4(v\eta  - M)v\eta \sin^2\left[(2p+1)\pi/2N)\right] . 
\eeq    
With the antiperiodic fermionic boundary condition   
the lowest fermionic mode in the continuum   
has a mass $M^2_0 =  M^2 + \pi^2/R^2$.   
In the small $M \ll \eta v$ limit  we   
obtain $M^2_0 =  M^2 + \eta^2v^2\pi^2/N^2$   
and, using $vR=N$ from the   
bosonic case the matching of the low energy modes     
to the continuum requires a      
choice of $\eta=1$. More generally, for   
any $M$ we require for matching to   
the continuum    
\beq     
-\widetilde{M}\eta/v = (\eta v - M)\eta/v = 1  . 
\eeq      
For $M >\eta v$ the sign of $\eta$ flips.    
      
The spectrum of  eq.~(\ref{wilsonterm}) is now faithful     
to the continuum limit (see Fig.~\ref{wmass2}).  Consider $N$-odd:     
We have the levels $p=[0,-1]$,  $p=[1,-2]$, ...      
and the previous     
zero-mode becomes now the      
high-mass singlet $p=[(N-1)/2]$.     
For $N$-even we have the same tower of     
low mass states terminating     
at the doubled highest mass levels $p= [N/2-1, N/2]$.      
We do not      
really care about the highest energy states     
since we cut the theory of for some level     
$p \ll N/2$.     
There is now only one basin of levels in     
both cases.     
     
The fermion doubling problem underlies   
difficulties for the faithful lattice implementation of SUSY. 
With the naive latticization, fermions are doubled, while 
bosons are not, thus breaking SUSY.    
Of course, this traces to the fact that the Lorentz   
group, $O(4,1)$ has been replaced by the group   
$O(3,1)\times Z_N$.  Thus, at the most fundamental level,   
the supersymmetric grading of the Lorentz group $O(4,1)$   
is not expected to be implemented in the lattice approximation. 
Deconstructions of SUSY theories to date treat the 
SUSY hopping terms as parts of a superpotential 
\cite{susy}. However, 
they should properly emerge from Kahler potentials 
in the higher dimensional theory, 
and it is likely that other terms or constraints will arise 
that are required for anomaly matching, etc.   
There is also the issue of chirality, such as 
the localization of domain wall fermions \cite{jackiw},  
which  is an interesting story in itself.  This has a nice   
realization in terms of the lattice construction: 
the chiral 
fermion shows up as a dislocation 
in the lattice hopping terms \cite{wang0} (see also 
Hill, He and Tait in ref.(\cite{dob})).   
We will return to these and other issues elsewhere \cite{jingx}.

\section{The Effective Potential}            
   
We now calculate the Coleman-Weinberg potential for the      
WLPNGB field $\chi$ by integrating     
out the fermions.  This amounts to computing     
the fermionic determinant for the Lagrangian     
eq.~(\ref{latferm}) in the classical     
background $\phi$:     
\beq     
V = +i\hbar \ln \left[ \det({i\slash{D} - M(\chi)}) \right] .    
\label{diracdet}     
\eeq     
To leading order we neglect the $1+3$ vector     
potentials, and replace each $\Phi_n$ by their   
common dependence upon the zero mode, $\chi_0$:     
\beq     
\Phi_n \rightarrow \Phi \equiv (v/\sqrt{2}g)\exp(ig \chi_0/\sqrt{N}v) ,   
\eeq    
where we have used the normalization conventions of eq.~(\ref{phi}).    
The Dirac action with the Wilson term is now:     
\beq     
\sum_{n=1}^N\int d^4x \; \left[ \bar{\tilde\Psi}_n(i\slash{\partial}     
 - \widetilde{M} )\tilde\Psi_n -      
 \eta v \bar{\tilde\Psi}_{nL}\exp(i g\chi_0/\sqrt{N}v) \tilde\Psi_{(n+1)R}  
 + h.c.\right]     
 \label{latferm5}     
\eeq     
Going to the $p$-basis we obtain:   
\bea    
& & \sum_{p=-J}^{J+\delta}\int d^4x \;\left[ \bar{\Psi}_p(   
i\slash{\partial}     
 -\widetilde{M}-\eta v \cos\left[\Omega_p \right])\Psi_p   
 -      
i\eta v \sin\left[\Omega_p \right]\bar{\Psi}_p \gamma_5 \Psi_{p}   
\right]      
 \label{latferm10}     
\eea     
where:   
\beq   
\Omega_p\equiv\frac{(2p+1)\pi v +g\sqrt{N}\chi_0}{Nv}  . 
\eeq   
This shifts the mass of each mode:   
\beq     
\label{wilsonterm2}     
M_p^2 = M^2 + 4(v\eta  - M)v\eta \sin^2   
\left[\half\Omega_p\right]     
= \widetilde{M}^2 + \eta^2v^2 + 2\widetilde{M}v\eta \cos(\Omega_p)  . 
\eeq    
The mass term is periodic under:   
\beq   
\chi_0 \rightarrow \chi_0 + 2\pi \sqrt{N}v/g   
\eeq   
which corresponds, with our normalization, to the modular invariance   
of the Wilson line under gauge transformations.   
   
The functional integral over fermions yields the Dirac   
determinant:   
\beq     
Z = \det({i\slash{D} - M(\chi_0)}) = \Pi_{k,p} \det(\slash{k} - M_p(\chi_0)),
\label{diracdet2}     
\eeq    
where the second expression contains a product   
over all 4-momenta and the discretized   
5th momentum, and  the determinant   
runs of the all four Dirac spin states for a given $(k,p)$.   
This latter expression applies when the mass matrix is   
diagonal. $Z$ is symmetric under an overall change in    
the signs of the $\gamma_\mu$, hence we can write:   
\beq     
Z^2 = \Pi_{k,p} [-k^2 + M_p^2(\chi_0)]^4   
\label{diracdet3}     
\eeq   
and the effective potential is:   
 \beq     
V= i\ln Z = 2 i \int \frac{d^4k}{(2\pi)^4} \sum_p \ln (-k^2 +   
M_p^2(\chi_0)) .   
\label{diracdet4}     
\eeq   
Here, and subsequently, we discard additive (non-$\chi_0$ 
dependent) constants. 
The $k$-integrals can be performed with a Wick-rotation 
and we use a Euclidean cut-off,  
$\Lambda$,   
obtaining to order $1/\Lambda^2$:   
\beq   
\label{result}   
V = -\frac{1}{16\pi^2} \sum_p 
  \left[ \Lambda^4\left(\ln \Lambda^4  -\half\right)   
+ 2M_p^2\Lambda^2 - M_p^4\ln\left(\frac{\Lambda^2}{M_p^2}\right) -   
\half M_p^4  \right] .   
\eeq    
Under $Z_N$ transformations, $\tilde\Psi_n\rightarrow\tilde\Psi_{n+m}$,    
$\chi_0$ is invariant,     
hence the induced     
potential for $\chi_0$ is invariant under $Z_N$.    
For $N\geq 3$ the explicit dependence upon $\chi_0$    
in $V$ is   
finite.  Therefore, the only $Z_N$ invariants that   
can arise for $N\geq 3$ in the sum involve the term 
with logarithm.   
Hence, $V$ can be written in a form, summed on $p$ for $N\geq 3$:   
\beq   
\label{result2}   
V   
% = -\frac{1}{64\pi^2} \sum_p  \left[    
%\left( \widetilde{M}^2 + \eta^2v^2 + 2\widetilde{M}v\eta\cos(\Omega_p )  
%\right)^2 \ln \left( 1 +    
%\frac{2\widetilde{M} v\eta \cos(\Omega_p)}{\widetilde{M}^2+    
%\eta^2v^2 }\right) \right]   
%\nonumber \\   
 =  -\frac{v^4}{16\pi^2} \sum_p  \left[    
\left( \eta^{-2} + \eta^2 - 2\cos(\Omega_p ) \right)^2    
\ln \left( 1 -    
\frac{2\cos(\Omega_p)}{\eta^{-2} +    
\eta^2 }\right) \right]  , 
\eeq   
where we have used the matching condition, $ -\widetilde{M}\eta = v$.  
This does not yet display the full $Z_N$ suppression,   
and we must therefore expand the logarithm
(the $n<3$ terms average to zero upon
summing over $p$):   
\beq   
\label{result3}   
V   
%=  -\frac{v^4}{64\pi^2} \sum_p \sum_{n=1}^\infty \left[    
%(\eta^{-2} + \eta^2- \left(\exp   
%\left[i\Omega_p \right]  + h.c.\right))^2    
%\left(\frac{(\exp   
%\left[i\Omega_p \right]  + h.c.)^n}{n(\eta^{-2} + \eta^2)^{n}}\right) \right] 
%\nonumber \\   
 = \frac{v^4}{8\pi^2} \sum_p \sum_{n=3}^\infty \left[     
\frac{1}{n(n-1)(n-2)} \left(\frac{(\exp   
\left[i\Omega_p \right]  + h.c.)^{n}}{(\eta^{-2} + \eta^2)^{n-2}}\right)    
\right] ,    
\eeq   
where additive constants that have no $\chi_0$ dependence   
and terms unimportant for $N\geq 3$ are dropped.   
   
Let us consider the natural 
limit in which $M$ is small compared to $v$ 
(but $M$ can be arbitrary compared to $1/R$ 
since $v=N/R$).  We have $\widetilde{M} = M-\eta v$   
and the matching condition, $ -\widetilde{M}\eta = v$. Hence, in   
the limit $M/v\rightarrow 0$ we have $\eta \rightarrow \pm 1$ and   
$\widetilde{M} \rightarrow \mp v$:   
\bea   
\label{result31}   
V & = &    
\frac{v^4}{2\pi^2} \sum_p \sum_{n=3}^\infty    
\frac{1}{n(n-1)(n-2)} \cos^n   
\left[\Omega_p \right] .   
\eea   
Expanding $\cos^n(\Omega) = (e^{i\Omega}/2 + e^{-i\Omega}/2)^n$ gives:  
\beq   
\label{result31a}   
V  
%& = &    
%-\frac{v^4}{8\pi^2} \sum_p \sum_{n=3}^\infty   
%\frac{2^{-n}}{n(n-1)(n-2)}\sum_{m=0}^n     
%\left(\begin{array}{c}   
%n\\   
%m   
%\end{array}\right)   
%\exp\left[i (n-2m) \Omega_p \right]\\   
% 
= \frac{v^4}{\pi^2} \sum_p \sum_{n=3}^\infty   
\frac{2^{-n}}{n(n-1)(n-2)}\sum_{m=0}^{[n/2]}     
\left(\begin{array}{c}   
n\\   
m   
\end{array}\right)   
\cos\left[(n-2m) \Omega_p \right].   
\eeq   
The only $\chi_0$ dependent terms that   
can survive the sum over $p$ in the expansion involve  
\beq   
\left(\exp   
\left[i\Omega_p\right]\right)^{qN}\qquad \qquad    
\eeq   
for integer $q$, a consequence of the $Z_N$ invariance.  Such terms  
reduce the sum on $p$ to an overall factor of $(-1)^qN$. 
(Note: with periodic fermion boundary conditions the sum on $p$ gives $N$ 
and the overall sign of the potential is flipped; this is  
consistent with the Wilson line redefinition of the fermionic 
boundary condition.)  Therefore,  
the sum over $p$ will only give contributions for $n-2m = qN$ for  
integer $q$:  
\bea   
\label{result31b}   
V & = &    
\frac{v^4N}{\pi^2} \sum_{q=1}^\infty \sum_{n=qN}^\infty \frac{(-1)^q    
2^{-n}}{n(n-1)(n-2)} \left(\begin{array}{c}   
n\\   
(n-qN)/2   
\end{array}\right)   
\cos\left[q \sqrt{N}g\chi_0/v \right], \\   
 & = &    
\frac{v^4 N}{\pi^2} \sum_{q=1}^\infty (-1)^q   
\cos\left[q \sqrt{N}g\chi_0/v \right]    
\sum_{k=0}^\infty  
2^{- q N - 2 k}\frac{(q N + 2 k - 3)!}{k! (q N+ k)!},  
\eea   
where $2k=n-qN$.  Using an exact result: 
\bea   
\sum_{k=0}^\infty 2^{-a -2k} \frac{(a+2k-b)!}{k!(a+k)!}   
= \frac{2^{b-1}\Gamma[1+a-b] \Gamma[b-1/2]}{\sqrt{\pi}\, \Gamma[a+b]}  
\eea   
we get:  
\bea   
\label{result31c}   
V & = &    
\frac{3 v^4}{\pi^2} \sum_{q=1}^\infty    
\frac{(-1)^q \cos(q \sqrt{N}g\chi_0/v)}   
{q (q^2N^2-1)(q^2N^2-4)}.   
\eea   
Hence, for large $N$ the leading term in the above series is:   
\bea   
\label{result32}   
V & \approx &    
-\frac{3v^4}{N^4\pi^2}      
\cos   
\left(g\sqrt{N}\chi_0/v\right)\\   
&=& -\frac{3}{\pi^2 R^4}      
\cos   
\left(\chi_0 /f_\chi\right),  \qquad f_\chi = 1/ \widetilde{g} R, 
\eea   
where   
$\widetilde{g} = g/\sqrt{N}$ is the low energy value of the coupling   
constant.  
 
From this analysis 
we obtain the 
{\em decay constant} of the $\chi_0$ 
field, $f_\chi= 1/ \widetilde{g} R$.  
 We have verified that this result 
is precise for large-$N$ 
by performing the sums numerically. 
Remarkably, the effective 
potential is finite for $N\geq 3$. Moreover, it 
has no cut-off dependence upon $N$, once we reexpress 
the parameters in terms of the low energy 
variables, $\tilde{g}$ and $R$.   
 
The minima of the potential occur for $\chi_0/f_\chi = 2n\pi$. 
This corresponds to the fermion acquiring {\em 
antiperiodic boundary conditions}. 
(had we computed the potential with fermions 
having periodic boundary conditions,  
the overall sign of $V$ would have flipped, and the minima 
would occur at $\chi_0/f_\chi = (2n+1)\pi$, hence the two 
computations are consistent.) 
The mass of the 
$\chi_0$ field is obtained by expanding about 
a minimum: 
\beq 
m_\chi^2 = \frac{3\tilde{g}^2}{\pi^2 R^2} = \frac{12}{\pi R^2}\tilde{\alpha}. 
\eeq   
This is similar to the result for a nonabelian  
gauge theory WLPNGB, $\propto \tilde{\alpha}/R^2$ 
\cite{ACG}.  
 
We also consider the limit, $\eta\ll1$, 
($\eta\gg1$ can be gotten by flipping the sign of the $\eta$ 
exponent in eq.~(\ref{smalleta})) 
which is a fermion with the
hopping links suppressed (strongly coupled). Eq.~(\ref{result3}) 
yields a leading $Z_N$ invariant term: 
\bea
V &=& 
\frac{v^4}{4\pi^2} \sum_{q=1}^\infty    
(-1)^q \frac{(\eta^2 + \eta^{-2})^{2 - N q}}{q (q N-1)(q N-2)}
\cos(q \sqrt{N}g\chi_0/v)\nonumber  \\
&&\phantom{\frac{v^4}{16\pi^2} \sum_{q=1}^\infty}
\times\;{}_2F_1[qN/2-1,(qN-1)/2;
qN+1;4/(\eta^2 + \eta^{-2})^2].
\eea
For $\eta\ll1$, the hypergeometric function goes to one, and the
potential reduces to:
\beq  
V \approx -\frac{v^4}{4\pi^2}\frac{\eta^{2(N-2)}}{N^2} 
\cos   
\left(g\sqrt{N}\chi_0/v\right) 
= -\frac{4e^{-(8\pi/\tilde\alpha)|\ln(\eta)|}}{R^4\tilde{\alpha}^2} 
\cos   
\left(\chi_0/f_\chi\right). 
\label{smalleta}
\eeq 
The potential is now $N$--dependent, and corresponds to the 
result for a generalized $Z_N$ theory (i.e., 
a ``theory space'' model). 
In the second expression we have swapped the explicit $N$ dependence 
for the ratio of the unitarity bound, $\sim 4\pi$, 
to the low energy coupling $\tilde{\alpha}=\tilde{g}^2/4\pi$,
i.e., setting $N = 4\pi/\tilde{\alpha}$. The expression 
is mysteriously remniscent of a dilute gas-approximation 
instanton potential.  
We see that the small (or large) $\eta$ limit  
produces an exponentially 
suppressed effective potential, and 
therefore an ultra-low-mass WLPNGB. 
In the parallel paper we show that this 
ultra-low-mass WLPNGB is immune to Planck scale 
breaking effects, and is a natural candidate for 
an axion, or a quintessence field \cite{adampngb}.

\section{Axial Anomaly}

A final issue of importance is that of anomalies.  The $U(1)^N$ 
theory in $1+3$ contains $N$ NGB's associated with the hopping 
terms.  Eq.~(\ref{latfermw2}) with the $\Phi_n= (v/\sqrt{2}g) 
\exp(ig\tilde{\chi}_n/v)$ displayed takes 
the form: 
 \bea     
& & \sum_{n=1}^N\int d^4x \; \left[ \bar{\tilde\Psi}_{n}(i\slash{\partial} 
-g\slash{A}_n)\tilde\Psi_{n}   
 - \widetilde{M}( \bar{\tilde\Psi}_{nL}\tilde\Psi_{nR} + h.c.) \right]  \\   
 & &    
+\sum_{n=1}^N\int d^4x \; \left[     
\left(-\eta v \bar{\tilde\Psi}_{nL}e^{ig\tilde{\chi}_n/v}\tilde\Psi_{(n+1)R}     
+ h.c.\right) + \half v^2(\partial\tilde{\chi}_n/v  
+ \tilde{A}_n  - \tilde{A}_{n+1})^2 \right]. 
 \label{anom1}  \nonumber  
\eea     
Do the chiral fields, $\tilde{\chi}_n$ develop anomalous couplings 
to the gauge fields? 
 
First, we must define the anomalies in the fundamental 
currents in a manner that is consistent with $U(1)^N$ and 
$Z_N$. The theory has $2N$ relevant 
currents, and each vectorial $U(1)$ 
must be anomaly free.  Therefore we define: 
\beq 
\partial^\mu \tilde\Psi_n\gamma_\mu\tilde\Psi_{nL} = 
-\frac{g^2}{32\pi^2}F_{n\mu\nu}\;{}^\star {F}^{\mu\nu}_n 
\qquad  
\partial^\mu \tilde\Psi_n\gamma_\mu\tilde\Psi_{nR} = 
\frac{g^2}{32\pi^2}F_{n\mu\nu}\; {}^\star {F}^{\mu\nu}_n 
\eeq 
where ${}^\star{F}_{n\mu\nu}= 
\half\epsilon_{\mu\nu\rho\sigma}{F}^{\mu\nu}_n$ 
and $F_{n\mu\nu}= \partial_\mu \tilde{A}_{n\nu}- 
\partial_\nu \tilde{A}_{n\mu}$. 
 
The key observation 
presently is that we can perform a sequence of $N$   
{\em vectorial} redefinitions of 
the fermions $\tilde\Psi_n$ and gauge transformations 
of the $\tilde{A}_{n\mu}$ which remove all but $\chi_0$ from the mass terms. 
Because these are {\em vectorial} transformations on the 
fermions, 
they have no associated anomalies.   
Let: 
\bea 
\tilde\Psi_n & \rightarrow &  
\exp\left[-ig\sum_{k=1}^{n-1}\tilde{\chi}_{k}/v +  
ig(n-1)\chi_0/\sqrt{N}v\right] \tilde\Psi_n, 
\\ \nonumber 
\tilde{A}_{n\mu} & \rightarrow & \tilde{A}_{n\mu}+\sum_{k=1}^{n-1}\partial_\mu\tilde{\chi}_{k}/v 
- (n-1)\partial_\mu\chi_0/\sqrt{N}v, 
\eea 
where, recall, $\chi_0 = \sum_{n=1}^N\tilde{\chi}_{n}/\sqrt{N}$. 
 
These transformations remove all of the $\tilde\chi_n$ from 
all hopping terms of eq.~(\ref{anom1}), except for a residual 
$\chi_0$ factor, and bring the vector potentials into 
the ``unitary gauge:'' 
\bea     
& & \sum_{n=1}^N\int d^4x \;  \bar{\tilde\Psi}_{n}(i\slash{\partial} 
-g\slash{A}_n)\tilde\Psi_{n}   
 - \widetilde{M}( \bar{\tilde\Psi}_{nL}\tilde\Psi_{nR} + h.c.) \\   
 & &    
-\sum_{n=1}^N\int d^4x \;     
\left[\eta v 
\bar{\tilde\Psi}_{nL}e^{ig\chi_0/\sqrt{N}v} 
\tilde\Psi_{(n+1)R}     
+ h.c.\right]    + \half(\partial\chi_0)^2  
+ \half v^2(\tilde{A}_n  - \tilde{A}_{n+1})^2. \nonumber  
 \label{anom2}     
\eea     
Hence we learn that {\em the only chiral anomalies 
involve the $\chi_0$ zero mode}.    
The $\chi_0$ zero mode can now be removed from 
the hopping terms, but 
this necessitates a {\em chiral} redefinition 
of the fermions, and it leads to a 
Wess-Zumino term. 
 
The purely anomalous contribution to the 
effective Lagrangian occurs when there is 
a classical chiral symmetry.  
The Wilson term, which behaves like a bosonic kinetic 
term for the fermions, 
violates chirality. 
The model Lagrangian possesses the classical 
chiral symmetry when  
$\widetilde{M}= M-\eta v = 0 $.  
 
In this 
limit we can redefine the fermions under a 
chiral transformation: 
\bea 
\tilde\Psi_{nL} & \rightarrow &  
\exp\left[ig\chi_0/2\sqrt{N}v\right]\tilde\Psi_{nL}, 
\\ \nonumber 
\tilde\Psi_{nR} & \rightarrow &  
\exp\left[-ig\chi_0/2\sqrt{N}v\right]\tilde\Psi_{nR}. 
\eea 
This produces, however, the Wess-Zumino term 
which is then added to the Lagrangian: 
\beq    
\frac{g\chi_0}{\sqrt{N} v}\sum_{n=1}^N 
\partial_\mu \overline{\tilde\Psi}_n\gamma^\mu\gamma^5\tilde\Psi_n 
= 
\frac{{g}^3}{16\pi^2 }   
\frac{\chi_0}{\sqrt{N} v} \sum_{n=1}^N  
F_{n\mu\nu}{}^\star{F}^{n,\mu\nu}.  
\eeq  
Now, we use the relationships to the low energy 
parameters, $v/N = 1/R$, and $\tilde{g}=g/\sqrt{N}$ 
and the Wess-Zumino term becomes: 
\beq    
\frac{\tilde{g}^2}{16\pi^2 }   
\frac{\chi_0}{f_\chi} \sum_{n=1}^N  
F_{n\mu\nu}{}^\star {F}^{n,\mu\nu}; \qquad f_\chi = \frac{1}{\tilde{g}R}.  
\eeq 
The decay constant of 
the WLPNGB is  
that obtained in the Coleman-Weinberg potential 
analysis of Section 3, and is given by the inverse  
of the product of the size $R$ of the extra dimension 
and the low energy coupling constant $\tilde{g}$. 
Thus the WLPNGB couples universally and anomalously  
to all KK modes, in analogy to the $\pi^0\rightarrow 2\gamma$  
coupling. 
 
When nonzero $\widetilde{M}$ is considered, we have 
explict breaking of the chiral symmetry and the coupling 
of $\chi_0$ to $F_{n\mu\nu} {}^\star {F}^{n,\mu\nu}$ is modified. 
For further discussion of this, in application to 
axion and quintessence physics, see \cite{adampngb}.  
%\newpage 
   
\section{Conclusions}            
       
The theory we have described, QED in $1+4$, has been  
periodically compactified and latticized 
to produce an equivalent $U(1)^N$ theory describing physics for 
a finite set of KK-modes, in the low energy 
limit $n\ll N$.   
 
In general a zero-mode $\chi_0$, occurs which 
may be interpreted as 
the $5$th component of the vector potential, or the  
Wilson line around the compact $5$th dimension. 
This possesses 
only discrete symmetries under gauge transformations, i.e., 
$\chi_0\rightarrow \chi_0 + 2\pi n v/\tilde{g}$, and the  
$U(1)$ action 
is generally broken to the $Z_N$ subgroup. 
In the effective $1+3$ Lagrangian $\chi_0$ appears as a  
pseudo-Nambu Goldstone Boson (WLPNGB).  In the  
absence of matter fields, the WLPNGB is massless. 
In general, however, the WLPNGB acquires a mass with a periodic 
potential, consistent with the discrete $Z_N$ invariance.   
 
We introduced matter fields into the latticized theory and 
encounter the Wilson flavor doubling problem. 
We have seen that the Wilson doubling 
has a remarkable diagrammatic 
interpretation, as two independent zig-zag chiral 
hopping threads through the lattice when the Dirac 
mass $M=0$. 
One intriguing model building possibility, which 
we have not presently explored, would be to  
allow the Wilson doubling 
phenomenon to be the {\em fundamental} origin of flavor.  This requires,  
however, dealing with the issues of imbedding the structure 
into the Standard Model and its associated flavor-chiral 
gauge structure.  It is not clear that sensible models exist, 
but we are presently investigating this possibility \cite{jingx}.

If one desires, however, a simple lattice description 
that is faithful to the continuum theory, one must eliminate the  
flavor doubling. This necessitates including the 
Wilson term.  The Wilson term improved action is studied 
and utilized in our present analysis.  One obvious consequence of 
the fermionic flavor doubling problem is 
that faithfully representing latticized SUSY theories 
is subtle, and may be problematic \cite{jingx}. To our knowledge, 
deconstructed SUSY models discussed in the literature to 
date, \cite{susy}, have not addressed this problem. The SUSY kinetic 
terms implemented in these analyses are typically superpotentials,  
and are not  deconstructed Kahler potentials.  This problem is 
fundamental to the deconstruction approach since the Poincare group 
is modified by the lattice, the extra dimensional continuous translational 
symmetries are replaced by $Z_N$.  
 
We obtain the Coleman-Weinberg potential for the WLPNGB, 
which is finite for large $N$.  
The finiteness is a consequence of the $Z_N$ symmetry. 
Such $Z_N$ finite potentials have been discussed 
previously, e.g.,  in the schizon model of Hill and Ross, 
\cite{ross}, which remarkably has the equivalent structure to 
the present scheme in the extra-dimension's 
momentum space.   Moreover, 
when recast in terms of the physical  
variables, $R$ (the size of the extra dimension) 
and $\tilde{g}$ (the low energy $3+1$ QED coupling 
constant)  all dependence upon the high energy 
scales, $N$, or $v$ ($v=1/a$ where $a$ is the lattice spacing), 
completely disappears in the Coleman-Weinberg potential. 
Therefore, the Coleman-Weinberg potential 
for the Wilson line 
is  reliably determined and is independent 
of the regulator scheme.  

When we calculate with antiperiodic (periodic)
fermions, we find that 
the Coleman-Weinberg potential is minimized for  
the special case $\chi_0=0$  ($\chi_0=\pi f_\chi$) 
(modulo the periodicity).  
Upon absorbing the Wilson line into 
the fermionic wave-function, this corresponds to the fermions 
always having dynamically
preferred {\em antiperiodic boundary conditions} in traversing the 
extra dimension.  
 
We study the anomaly structure of the deconstructed theory and 
find that only $\chi_0$ develops a Wess-Zumino term.  This  
WZ-term is the analogue of the $\pi\rightarrow 2\gamma$ anomaly 
in electrodynamics. In the present case the anomaly universally 
couples $\chi_0$ to all KK modes.  
  
By tweaking 
the parameters in the theory 
we are able to exponentially suppress the 
Coleman-Weinberg potential and  
generate $\chi_0$ as an ultra-low-mass spin-$0$ particle. 
This suggests a number of phenomenological applications.  
The effect of Planck-scale breaking  
of global symmetries is highly suppressed by the $Z_N$ symmetry. 
The construction of natural models of 
the axion and quintessence will be described elsewhere 
\cite{adampngb}.

%\noindent              
\section*{Acknowledgements}              
              
We wish to thank W. Bardeen for useful discussions, and 
Hsin Chia Cheng who corrected an overall 
sign error in our original result for the 
effective potential.              
Research by CTH and AKL was supported by the U.S.~Department of Energy    
Grant DE-AC02-76CH03000.              
         
% \newpage      

%\vspace*{1.0cm}              
%\newpage              
            
\frenchspacing              
%\vspace*{1cm}              
%\vspace*{1.0cm}              
%\noindent              
%\newpage              
              

\begin{thebibliography}{99}              
 
%\cite{Coleman:jx} 
\bibitem{cole} 
S.~Coleman and E.~Weinberg, 
%``Radiative Corrections As The Origin Of Spontaneous Symmetry Breaking,'' 
Phys.\ Rev.\ D {\bf 7}, 1888 (1973). 
%%CITATION = PHRVA,D7,1888;%% 
 
\bibitem{HPW}       
C.~T.~Hill, S.~Pokorski and J.~Wang,       
%``Gauge invariant effective Lagrangian for Kaluza-Klein modes,''       
Phys.\ Rev.\ D {\bf 64}, 105005 (2001).       
%%CITATION = HEP-TH 0104035;%%       
       
     
     
\bibitem{wang0}  
H.~C.~Cheng, C.~T.~Hill, S.~Pokorski and J.~Wang, 
%``The standard model in the latticized bulk,'' 
Phys.\ Rev.\ D {\bf 64}, 065007 (2001);
%%CITATION = HEP-TH 0104179;%% 
H.~C.~Cheng, C.~T.~Hill and J.~Wang, 
%``Dynamical electroweak breaking and latticized extra dimensions,'' 
Phys.\ Rev.\ D {\bf 64}, 095003 (2001). 
%%CITATION = HEP-PH 0105323;%% 
    
         
     
%\cite{Arkani-Hamed:2001ca}       
\bibitem{ACG}       
N.~Arkani-Hamed, A.~G.~Cohen and H.~Georgi,       
%``(De)constructing dimensions,''       
Phys.\ Rev.\ Lett.\  {\bf 86}, 4757 (2001);
%%CITATION = HEP-TH 0104005;%%       
 N.~Arkani-Hamed, A.~G.~Cohen and H.~Georgi, 
%``Electroweak symmetry breaking from dimensional deconstruction,'' 
Phys.\ Lett.\ B {\bf 513}, 232 (2001).    
       
%\cite{Georgi:au}    
\bibitem{georgi}    
H.~Georgi and A.~Pais,    
%``CP - Violation As A Quantum Effect,''    
Phys.\ Rev.\ D {\bf 10}, 1246 (1974).    
%%CITATION = PHRVA,D10,1246;%%    
    
\bibitem{ross}     
C.~T.~Hill and G.~G.~Ross,     
%``Models And New Phenomenological Implications Of A Class Of Pseudogoldstone Bosons,''     
Nucl.\ Phys.\ B {\bf 311}, 253 (1988);      
%%CITATION = NUPHA,B311,253;%%      
C.~T.~Hill and G.~G.~Ross,     
%``Pseudogoldstone Bosons And New Macroscopic Forces,''     
Phys.\ Lett.\ B {\bf 203}, 125 (1988).     
%%CITATION = PHLTA,B203,125;%%    
%\cite{Hill:1988bu}  
 
 
      
%\cite{Hill:1989vm}     
\bibitem{hillc}     
C.~T.~Hill, D.~N.~Schramm and J.~N.~Fry,     
%``Cosmological Structure Formation From Soft Topological Defects,''     
Comments Nucl.\ Part.\ Phys.\  {\bf 19}, 25 (1989);     
%%CITATION = CNPPA,19,25;%%     
C.~T.~Hill, P.~J.~Steinhardt and M.~S.~Turner,     
%``Can Oscillating Physics Explain An Apparently Periodic Universe?,''     
Phys.\ Lett.\ B {\bf 252}, 343 (1990);      
%%CITATION = PHLTA,B252,343;%%     
J.~A.~Frieman, C.~T.~Hill and R.~Watkins,     
%``Late time cosmological phase transitions. 1. Particle physics models and cosmic evolution,''     
Phys.\ Rev.\ D {\bf 46}, 1226 (1992);  
%%CITATION = PHRVA,D46,1226;%%
J.~A.~Frieman, C.~T.~Hill, A.~Stebbins and I.~Waga, 
%``Cosmology with ultralight pseudo Nambu-Goldstone bosons,'' 
Phys.\ Rev.\ Lett.\  {\bf 75}, 2077 (1995). 
%%CITATION = ASTRO-PH 9505060;%%  
 
\bibitem{kolb}  
A.~K.~Gupta, C.~T.~Hill, R.~Holman and E.~W.~Kolb,     
%``Statistical mechanics of soft boson phase transitions,''     
Phys.\ Rev.\ D {\bf 45}, 441 (1992).     
%%CITATION = PHRVA,D45,441;%%      
  
\bibitem{adampngb}
C.~T.~Hill and A.~K.~Leibovich,
%``Natural theories of ultra-low mass PNGB's: Axions and quintessence,''
arXiv:hep-ph/0205237.
%%CITATION = HEP-PH 0205237;%%


  
       
%\cite{Hill:2000mu}       
     
     
%\cite{Dimopoulos:2002jf}     
\bibitem{dim} 
M.~A.~Luty and R.~Sundrum, 
%``Supersymmetry breaking and composite extra dimensions,'' 
Phys.\ Rev.\ D {\bf 65}, 066004 (2002);
%%CITATION = HEP-TH 0105137;%%     
S.~Dimopoulos, D.~E.~Kaplan and N.~Weiner,     
%``Electroweak Unification into a Five-Dimensional SU(3) at a TeV,''     
arXiv:hep-ph/0202136;
%%CITATION = HEP-PH 0202136;%%      
N.~Weiner, 
%``Unification without unification,'' 
arXiv:hep-ph/0106097; 
%%CITATION = HEP-PH 0106097;%% 
J.~Dai and X.~C.~Song, 
%``Spontaneous symmetry broken condition in (de)constructing dimensions  from noncommutative geometry,'' 
arXiv:hep-ph/0105280; 
%%CITATION = HEP-PH 0105280;%% 
 M.~Bander, 
%``Gravity in dynamically generated dimensions,'' 
Phys.\ Rev.\ D {\bf 64}, 105021 (2001); 
%%CITATION = HEP-TH 0107130;%%   
N.~Arkani-Hamed, A.~G.~Cohen, D.~B.~Kaplan, A.~Karch and L.~Motl, 
%``Deconstructing (2,0) and little string theories,'' 
arXiv:hep-th/0110146. 
%%CITATION = HEP-TH 0110146;%% 
           
%\cite{Dob}     
\bibitem{dob}   
C.~T.~Hill,     
%``Topcolor: Top quark condensation      
%in a gauge extension of the standard model,''     
Phys.\ Lett.\ B {\bf 266}, 419 (1991);     
%%CITATION = PHLTA,B266,419;%%;     
C.~T.~Hill,     
%``Topcolor assisted technicolor,''     
Phys.\ Lett.\ B {\bf 345}, 483 (1995);     
%%CITATION = HEP-PH 9411426;%%
B.~A.~Dobrescu and C.~T.~Hill,     
%``Electroweak symmetry breaking via top condensation seesaw,''     
Phys.\ Rev.\ Lett.\  {\bf 81}, 2634 (1998);    
%%CITATION = HEP-PH 9712319;%%     
R.~S.~Chivukula, B.~A.~Dobrescu, H.~Georgi and C.~T.~Hill,     
%``Top quark seesaw theory of electroweak symmetry breaking,''     
Phys.\ Rev.\ D {\bf 59}, 075003 (1999);     
%%CITATION = HEP-PH 9809470;%%     
H.~J.~He, C.~T.~Hill and T.~M.~Tait, 
%``Top quark seesaw, vacuum structure and electroweak precision  constraints,''
Phys.\ Rev.\ D {\bf 65}, 055006 (2002). 
 
\bibitem{susy} 
C.~Csaki, J.~Erlich, C.~Grojean and G.~D.~Kribs, 
%``4D constructions of supersymmetric extra dimensions and gaugino  mediation,'' 
Phys.\ Rev.\ D {\bf 65}, 015003 (2002);
%%CITATION = HEP-PH 0106044;%% 
H.~C.~Cheng, D.~E.~Kaplan, M.~Schmaltz and W.~Skiba, 
%``Deconstructing gaugino mediation,'' 
Phys.\ Lett.\ B {\bf 515}, 395 (2001); 
%%CITATION = HEP-PH 0106098;%% 
C.~Csaki, G.~D.~Kribs and J.~Terning, 
%``4D models of Scherk-Schwarz GUT breaking via deconstruction,'' 
Phys.\ Rev.\ D {\bf 65}, 015004 (2002);
%%CITATION = HEP-PH 0107266;%% 
H.~C.~Cheng, K.~T.~Matchev and J.~Wang, 
%``GUT breaking on the lattice,'' 
Phys.\ Lett.\ B {\bf 521}, 308 (2001); 
%%CITATION = HEP-PH 0107268;%% 
N.~Arkani-Hamed, A.~G.~Cohen and H.~Georgi,
%``Twisted supersymmetry and the topology of theory space,''
arXiv:hep-th/0109082;
%%CITATION = HEP-TH 0109082;%%
T.~Kobayashi, N.~Maru and K.~Yoshioka, 
%``4D construction of bulk supersymmetry breaking,'' 
arXiv:hep-ph/0110117;
%%CITATION = HEP-PH 0110117;%% 
D.~Cremades, L.~E.~Ibanez and F.~Marchesano, 
%``SUSY quivers, intersecting branes and the modest hierarchy problem,'' 
arXiv:hep-th/0201205;
%%CITATION = HEP-TH 0201205;%% 
A.~Falkowski, C.~Grojean and S.~Pokorski, 
%``Soft electroweak breaking from hard supersymmetry breaking,'' 
arXiv:hep-ph/0203033;
%%CITATION = HEP-PH 0203033;%% 
Z.~Chacko, E.~Katz and E.~Perazzi, 
%``Yukawa deflected gauge mediation in four dimensions,'' 
arXiv:hep-ph/0203080;
%%CITATION = HEP-PH 0203080;%% 
P.~Brax, A.~Falkowski, Z.~Lalak and S.~Pokorski, 
%``Custodial supersymmetry in non-supersymmetric quiver theories,'' 
arXiv:hep-th/0204195. 
%%CITATION = HEP-TH 0204195;%% 
 
\bibitem{chris}   
C.~T.~Hill and P.~Ramond,   
%``Topology in the bulk: Gauge field solitons in extra dimensions,''   
Nucl.\ Phys.\ B {\bf 596} (2001) 243;
C.~T.~Hill,   
%``Topological solitons from deconstructed extra dimensions,''   
Phys.\ Rev.\ Lett.\  {\bf 88}, 041601 (2002);   
%%CITATION = HEP-TH 0109068;%%    
%%CITATION = HEP-TH 0007221;%%    
W.~Skiba and D.~Smith, 
%``Localized fermions and anomaly inflow via deconstruction,'' 
Phys.\ Rev.\ D {\bf 65}, 095002 (2002). 
%%CITATION = HEP-PH 0201056;%%  
C.~Csaki, J.~Erlich, V.~V.~Khoze, E.~Poppitz, Y.~Shadmi and Y.~Shirman, 
%``Exact results in 5D from instantons and deconstruction,'' 
Phys.\ Rev.\ D {\bf 65}, 085033 (2002); 
%%CITATION = HEP-TH 0110188;%% 
E.~Poppitz and Y.~Shirman, 
%``The strength of small-instanton amplitudes in gauge theories with  compact extra dimensions,'' 
arXiv:hep-th/0204075. 
 
     
 
 
 
 
 
   
\bibitem{kron}  
A.~S.~Kronfeld, 
%``Lattice QCD,'' 
FERMILAB-CONF-92-040-T 
{\it Introductory lectures given at TASI  
Summer School, Perspectives in the Standard Model,  
Boulder, CO, Jun 2-28, 1991}. 
  
 
 
\bibitem{jackiw}  
R.~Jackiw and C.~Rebbi,              
%``Solitons With Fermion Number 1/2,''             
Phys.\ Rev.\ D {\bf 13}, 3398 (1976).\\              
%%CITATION = PHRVA,D13,3398;%%              
D.~B.~Kaplan,             
%``A Method for simulating chiral fermions on the lattice,''             
Phys.\ Lett.\ B {\bf 288}, 342 (1992)             
hep-lat/9206013.   \\           
%%CITATION = HEP-LAT 9206013;%%              
N.~Arkani-Hamed and M.~Schmaltz,			              
%``Hierarchies without symmetries from extra dimensions,'              
Phys.\ Rev.\ D {\bf 61}, 033005 (2000)			                
hep-ph/9903417;  \\            
%%CITATION = HEP-PH 9903417;%%              
E.~A.~Mirabelli and M.~Schmaltz,              
%``Yukawa hierarchies from split fermions in extra dimensions,''              
Phys.\ Rev.\ D {\bf 61}, 113011 (2000)              
hep-ph/9912265;  \\            
%\bibitem{Kaplan:2000av}              
D.~E.~Kaplan and T.~M.~Tait,              
%``Supersymmetry breaking, fermion masses and a small extra dimension,''              
JHEP{\bf 0006}, 020 (2000)              
hep-ph/0004200;  \\            
%\cite{Dvali:2000ha}              
%\bibitem{dvali}              
G.~Dvali and M.~Shifman,              
%``Families as neighbors in extra dimension,''              
Phys.\ Lett.\ B {\bf 475}, 295 (2000)              
hep-ph/0001072.              
%%CITATION = HEP-PH 0001072;%%                
    
 
 
\bibitem{jingx} C.T.Hill,    
A. Leibovich, and J. Wang, work in progress.        
 
 
 
              
\end{thebibliography}
\end{document}